\documentclass[12pt,bezier]{article}
\usepackage{amsmath,amssymb,amsfonts,euscript}
\usepackage[usenames]{color}
\usepackage{amsmath,amssymb,amsfonts,epsf,euscript}
\usepackage[usenames]{color}
\usepackage{graphicx}
\newpage

\newcommand{\be}{\begin{equation}}
\newcommand{\ee}{\end{equation}}

\newtheorem{definition}{Definition}[section]

\title{{\bf\Large Hurst estimation of scale invariant processes with drift and stationary increments}}
\vspace{-1cm}
\author
{ N. Modarresi\thanks{\scriptsize
Department of Mathematics and computer science, Allameh Tabataba'i University, Tehran, Iran.
E-mail:  n.modarresi@atu.ac.ir(N. Modarresi).}
\and S. Rezakhah\thanks{\scriptsize
Faculty of Mathematics and Computer Science, Amirkabir University of
Technology, 424 Hafez Avenue, Tehran 15914, Iran. E-mail:
rezakhah@aut.ac.ir(S. Rezakhah).}
}
\date{}
\begin{document}

\maketitle

\begin{abstract}
The characteristic feature of the discrete  scale invariant (DSI) processes is the  invariance   of  their finite dimensional distributions  by   dilation for certain  scaling factor.
 DSI  process   with piecewise linear
 drift      and stationary increments inside  prescribed scale intervals     is introduced  and  studied. 
     To identify the structure of the process, first we determine the scale intervals, their linear drifts  and eliminate them. 
 Then  a new method for the estimation of the Hurst 
parameter of such DSI processes is presented  and  applied to some period of the Dow Jones indices.
This  method is based on  fixed number equally spaced  samples  inside successive scale intervals.   We also present some efficient 
method for estimating Hurst parameter of self-similar  processes with stationary increments. We compare the performance of this method with the celebrated FA, DFA and DMA  on the simulated data of fractional Brownian motion.\\ \\
{\it Mathematics Subject Classification MSC 2010:} 62L12; 60G22; 60G18.\\ \\
{\it keywords:} Discrete scale invariance; Hurst estimation; Fractional Brownian motion; Scale parameter.
\end{abstract}

\maketitle
\section{Introduction}
Scale invariance or self-similarity has been discovered, analyzed and exploited in many frameworks, such as natural images \cite{rud}, fluctuations of stock market \cite{con}, \cite{fei} and traffic modeling in broadband networks \cite{abr}, \cite{rao}.
These processes  are invariant in distribution under suitable scaling of time and space.  Discrete scale invariant (DSI) processes are invariant by dilation for certain preferred scaling factors
 or the observable  obeys scale invariance for specific choices of scale \cite{bor}, \cite{sor}, \cite{zho}.
In practice, the main object is detecting the scale invariant property and estimating the Hurst index {\color{blue}$H$} and  scale parameter {\color{blue}$\lambda$} of such processes \cite{b-1}. 
Estimation methods depend on several factors, e.g, the estimation technique, sample size, time scale, level shifts, correlation and data structure. 
Among all  estimation methods for scale invariant  and long-memory processes, 
the rescaled adjusted range or  $R/S$ statistic   and  semi-variogram are frequently used, see Beran \cite {b2}.   Balasis et al.  used the $R/S$  statistic in \cite{bal1} and  the wavelet spectral analysis in \cite{bal2} to estimate the Hurst exponents for self-similar time series
 originated from space physics applications.
Recently, Wang \cite{w1} presented a moving average method to estimate the Hurst exponent.
Vidacs and Virtamo obtained  maximum likelihood estimator of the Hurst parameter  based on some  geometric sampling   of the 
 fractional Brownian motion (fBm) traffic  \cite{v1}.
Some estimation methods of Hurst index are  based on variance-time, see  \cite{c1}, \cite{c2}.    In the econophysics, there are some celebrated methods for the estimation of Hurst parameters called  
 fluctuation analysis (FA)
\cite{p0},  detrended fluctuation analysis (DFA) 
 \cite{p00} and detrending moving average (DMA) which is described in  \cite{a0}.   

Regression analysis is a form of predictive modeling technique which allows to detect the trend of time series. We apply some piece-wise linear regression to detect drift to  the DSI processes.
Especially we determine such piece-wise linear drift  by applying  regression lines to the plots of the corresponding scale intervals  of DSI processes.\\
In real data scale invariant  behavior often occures with some linear drift.  To dtermine the structure of  such processes  one need to eliminate the drift first, and then estimate  the Hurst  parameter.
Brownian motion with drift is an example  of scale invariant processes with drift which  has lots of applications  in mathematical finance and  stock price modelings   l \cite{ros}.

Usually the  DSI behavior of the  processes are characterized by detecting  some regular behavior of the process inside  successive scale intervals which can be identified by fitting homologous  parabolas.   
In this paper we present some flexible sampling scheme which provide some fixed number equally spaced sample points in each scale interval. 
This sampling scheme provide a bases for our estimation method of Hurst parameter.
 Then  a DSI process with drift is introduced where by evaluating successive scale intervals, the evaluation and elimination of the linear piece-wise  drift is studied.  
Then a new innovative method for the estimation of  Hurst parameter of  DSI processes, having  stationary increments inside scale intervals,  is developed.
Finally  we present some heuristic  estimator of  the  Hurst parameter of self-similar processes with stationary increments (Hsssi).  
The performance of this estimator are examined by using simulated data. 
We show that our method is more efficient than the  FA, DFA and DMA methods.
 We compare our method using simulated samples of fractional Brownian motion (fBm) with drift and  different Hurst parameters and show that the mean square error (MSE) of our method is much less than the compared methods.\\
The paper is structured as follows. In section 2, we present our flexible discrete sampling scheme and provide intuition on some basic notions of the scale invariance and DSI processes in discrete parameter space.
Section 3 is devoted to introducing DSI processes with drift, piece-wise linear drift and its elimination. We also present our estimation methods  for scale and Hurst parameter of DSI processes while the increments are stationary inside scale intervals, and apply
 our estimation methods to some part of Dow Jones indices in section 3.  
 A heuristic method for estimating the Hurst parameter of self-similar  process with stationary increments is developed  and its performance is compared with the celebrated methods FA, DFA and DMA for simulated data of fBm in section 4.
 Conclusions are presented in section 5.

\section{Method of Flexible Discrete Sampling}
For our estimation method  some appropriate  sampling scheme is required.  Current authors \cite{m11}  considered geometric sampling  at points  $\alpha^k$, $k=1, 2, \ldots$  of   DSI processes  $\{X(t), t\in {\Bbb R^+}\}$  with scale $\lambda>1$, where $\alpha$ is determined by $\lambda=\alpha^T$ ,  and  $T\in {\Bbb N}$  is some predefined number of observations in the scale interval $[\lambda^n, \lambda^{n+1})$, $n=0, 1,2,\cdots$.    
Here we present some basic definitions and flexible sampling for DSI processes, see \cite{m3}.
A process $\{X(k),k\in {\check{T}}\}$ is called discrete time self similar (or  scale invariant) process with parameter space $\check{T}$,
where $\check{T}$ is any subset of countable distinct points of positive real numbers, if $\{X(k_2)\}\stackrel{d}{=}(\frac{k_2}{k_1})^H\{X(k_1)\}$ for any $k_1, k_2 \in \check{T}$, where $\stackrel{d}{=}$ denotes equality of finite dimensional distributions.
  The process is called discrete time DSI process  with scale {\color{blue}$\lambda>0$} and parameter space $\check{T}$, if for any $k_1, k_2=\lambda k_1 \in \check{T}$, the above equation holds  in distribution.  So 
 sampling of  $\{X(t), t\in {\Bbb R^+}\}$ at points $\alpha^{nT+k}, n\in {\Bbb Z}$, $k=0, 1, \ldots, T-1$, we have a discrete time scale invariant process with parameter space $\check{T}=\{\alpha^{nT+k}, n\in {\Bbb Z}\}$.
\\
A random process $\{X(k),k\in \check{T}\}$ is called self-similar ( scale invariant) in the wide sense with Hurst index $H>0$ and  parameter space $\check{T}$, if for all $k, k_1\in \check{T}$ and all  {\color{blue}$\lambda>0$}, where $\lambda k, \lambda k_1\in \check{T}$ we have that  $E[X^2(k)]<\infty$, $E[X(\lambda k)]=\lambda^HE[X(k)]$ and $E[X(\lambda k)X(\lambda k_1)]=\lambda^{2H}E[X(k)X(k_1)]$. If these properties hold for some $\lambda=\lambda_0>0 $ then the process is called wide sense DSI with parameter space  $\check{T}$,   see \cite{m11}.\\
Following Modarresi and Rezakhah  \cite{m3}  we consider discrete  flexible sampling   of DSI process with scale $\lambda>1$  by 
 choosing  arbitrary sample points of  in the first scale interval as 
  as $1\leqslant s_{1}< s_{2}< \ldots< s_{q}< \lambda$ and  sampling in the scale interval  $I_j=[\lambda^j, \lambda^{j+1})$, $j\in{\Bbb N}$, at points $\lambda^js_i$, $i=1, \ldots, q$. So by recalling sample points with $t_j=\lambda^{[\frac{j-1}{q}]}s_{j-[\frac{j-1}{q}]q}$ our sample space are being $\check{T}=\{t_j, j\in {\Bbb W}\}$.

\section{DSI processes with drift}
For the real data, DSI behavior often occurs in short periods. The Dst time series \cite{b-1} and stock market indices \cite{b0}, \cite{m3}, \cite{m4} and \cite{m114} are some examples of such situations. The change of drift is  another feature that specially occurs by the 
changes in growth  of stock markets. So  their simultaneous effect can not be ignored. 
  There are many examples in modeling the behavior of stock prices by Brownian motion with drift \cite{hul}.
Besides the regression modeling approaches,  Brownian motion with linear drift has drawn much attention in prognostics \cite{wang}.    Here we present the definition of Brownian motion with linear drift. 
\begin{definition}
The process $\{B_{\mu}(t), t\geqslant 0\}$ is a Brownian motion with drift coefficient $\mu$ and variance parameter $\sigma^2$ if $B_{\mu}(0)=0$, the process $\{B_{\mu}(t), t\geqslant 0\}$ has stationary and independent increments and $B_{\mu}(t)$ is normally distributed with mean $\mu t$ and variance $\sigma^2t$. Equivalently $B_{\mu}(t)=\sigma B(t)+\mu t$ is Brownian motion with drift, where{\color{blue}  the standard Brownian motion $B(t)$  is $H=1/2$ self-similar}.
\end{definition}
Now we present the definition of DSI processes with piece-wise linear drift, which we apply in subsection 3.3 to model 
some part of the Dow Jones indices . 
\begin{definition}
A process $\{X(t), t\in{\Bbb R}\}$ is DSI process with drift if it satisfies the relation $X(t)=Y(t)+g(t)$, $t\in{\Bbb R}$ where $Y(t)$ is a DSI process and $g(t)$ is a drift function.   We call the process $\{X(t)\}$ , DSI  with piece-wise linear drift  if the drift consist of  different line in successive scale intervals of the  DSI process as $g(t)=\sum_{k=1}^M (\alpha_k+\beta_k t) I_{B_k}(t)$, where $\alpha_k$ and $\beta_k$ are real numbers  and $B_k$ is the $k$-th scale interval for $k=1,\cdots M$.
\end{definition}

Following the idea of  
 the Brownian motion with drift, described in \cite{ros}, and simple Brownian motion as a DSI process, described in \cite{m11}, we present an example, which we call simple Brownian motion with drift as a flexible pattern for modeling more comprehensive processes with DSI behavior.
 
\noindent{\bf Example:}
Following Modarresi et al. \cite{m4} we call a process $X(t)$ a simple Brownian motion with drift $g(t)$,  the Hurst index $H>0$ and scale $\lambda>1$  if 
\vspace{-3mm}
$$X(t)=\sum_{n=1}^{M}\lambda^{n(H-\frac{1}{2})}I_{[\lambda^{n-1}, \lambda^{n})}(t)B(t)+g(t)$$
where $B(\cdot)$ is {\color{blue}the standard} Brownian motion, $I(\cdot)$ an indicator function.
The expected value of the process is 
$$E(X(t))=\lambda^{n(H-\frac{1}{2})}E(B(t))+g(t)=g(t).$$
 Also for $s\leqslant t$,  the covariance function of the process is determined as 
$$\mathrm{Cov}\big(X(t), X(s)\big)=\lambda^{(n+m)(H-\frac{1}{2})}\mathrm{Cov}\big(B(t)+g(t),
B(s)+g(s)\big)=\lambda^{(n+m)(H-\frac{1}{2})}s.  $$
 This by the fact that  $\mathrm{Cov}\big(B(t), B(s)\big)=\min\{t,s\}$.   
 One can easily verify that 
 $\{ X(t) \}$   is a  DSI process.   As in the real world the DSI behavior just appears locally in different processes,  one need to detect the DSI period first.

The rest of this section can be described as follows. In this section we introduce DSI processes with drift and give an example which has Markov property. 
In subsection 3.1  we study the detection and elimination of the piece-wise linear drift to the DSI processes. 
In subsection 3.2  we  introduce  some  innovative method for estimating the Hurst parameters of  DSI processes with stationary increments.  In subsection 3.3 we  apply  our estimation method  for estimating  the Hurst parameter of real data as some part   Dow Jones indices. 

\subsection{Elimination of the drift}
We consider the DSI process with piece-wise linear drift as $X(t)=Y(t)+g(t)$  in  some duration of time, say $[0, C]$,  where $\{ Y(t), t\in [0,C]\} $ is a DSI process and the drift  $g(t)$ is assumed to be 
 piece-wise  linear function of time $t$  as $g(t)=\sum_{i=1}^k  (\alpha_i+\beta_i)I_{B_i}(t) $ where $B_1, B_2, \cdots , B_n$  is some partition of  this  duration. 
For decomposition of $X(t)$ as above and detecting the DSI behavior of $Y(t)$ one need to estimate such piece-wise linear  drift and eliminate it from the main  process $X(t)$ first. For this, we need to detect the corresponding scale intervals of the main process  by fitting some parabola  as has been applied in \cite{b0} and \cite{m3}.  Then we   fit  separate regression lines to the  samples of successive scale intervals of main process $X(t)$. These regression lines are considered  as piece-wise linear drift to the process $\{ X(t), t \in [0,C] \}$.  
Then we eliminate the drift  by subtracting $g(t)$ from $X(t)$ to obtain the DSI process $Y(t)$. Eliminating this  drift Then one can estimate the Hurst parameter of the corresponding  DSI process $Y(t)$ by the following method.\\
 Using detected  scale intervals of $X(t)$ for the corresponding DSI process $Y(t)$,  we consider  some fixed number of equally spaced samples in each scale interval, 
say $q$.  By this and the assumption that the process have stationary increment property  inside each scale interval, we conclude that the increments arises by this sampling method inside each scale interval are identically distributed. 
Now we consider the following  procedure for the estimation of the parameters of the DSI process $Y(t)$.

\subsection{Estimation procedure }
Let $Y=\{Y(t), t\geqslant 0\}$ be a  DSI process with stationary increment property  inside each scale interval. This  cause the increments of equally spaced samples inside each  scale interval   to be identically distributed. Our 
estimation  method is presented by the following steps.\\ \\
1- The time interval that we study  the DSI  process $\{ Y(t) \}$  is considered as  $[0,C]$.\\ \\
2- Following the methods of Bartolozzi et al. \cite{b0} and Modarresi et al. \cite{m3} and \cite{m4}, we evaluate scale intervals $I_i=(a_i, a_{i+1}]$ by fitting appropriate parabola to the samples of the period $[0,C]$.  
   So the scale of the process can be estimated by
$$\hat{\lambda}=\frac{1}{M}\sum_{i=1}^{M}\frac{a_{i+1}-a_i}{a_i-a_{i-1}}$$
where $M$ is the number of the scale intervals.\\ \\
3-  We consider  $q$ equally spaced sample points in each scale interval so that  the sample points of  $k$-th  scale interval  are determined as  $t_{(k-1)q+i}=a_{k-1}+(i-1)d_k$, where $d_k=\frac{a_{k+1}-a_k}{q}$, $i=1, \ldots, q$ and $k= 1, \ldots, M$.  So  $q$ is the number of observations in each scale interval. Thus our parameter space is $\hat{\tau}=\{t_{(k-1)q+i}, i=1, \ldots, q, k= 1, \ldots, M\}$ and  $\{Y(t), t\in \hat{\tau}\}$ is a  DSI process with parameter space $\hat{\tau}$.\\ \\
4- Now we consider $\{U(t), t\in \hat{\tau}\}$ as the increment process, where $U(t_i)=Y(t_{i+1})-Y(t_{i})$,
 and  $S_{k}^2=\frac{1}{q}\sum_{i=1}^{q}(U(t_{(k-1)q+i})-\bar{U_k})^2$,
where $\bar{U_k}=\frac{1}{q}\sum_{i=1}^{q} U_{(k-1)q+i}$, the  sample variance of  increments  in the $k$-th scale interval, $k=1, 2, \ldots, M$.\\ \\
5- By the scale invariant property of the process $\{Y(t), t\in \hat{\tau}\}$ we have that $U(t_{(k-1)q+i})\stackrel{d}{=}\hat{\lambda}^H U(t_{(k-2)q+i})$ for $i=1, \ldots, q$, so $\sigma_k^2=\hat{\lambda}^{2H}\sigma_{k-1}^2$,    where $\sigma_k^2=\mbox{Var}\big(U(t_{(k-1)q+i})\big)$.  Estimating $\sigma^2_k$  by $S^2_k$, one can  evaluate the estimation of the Hurst parameter by $\hat{\lambda}^{2\hat{H}}=\frac{S^2_k}{S^2_{k-1}}$.
Denoting $\mu=\hat{\lambda}^{2H}$, we have $M-1$ estimate for $\mu$ as $\hat{\mu}_k=\frac{S^2_{k}}{S^2_{k-1}}$, $k=2, \ldots, M$ and the final estimation of $\mu$ is evaluated as the mean of these $\hat{\mu}_k$.\\ \\
This estimation method of the Hurst parameter is based on the first order of the increments. We also consider the second order difference of the increments inside scale intervals and evaluate the sample variance 
 of corresponding to $k$-th scale interval   by $S_{k,2}^2=\frac{1}{q-1}\sum_{i=1}^{q-1}(Z(t_{(k-1)(q-1)+i})-\bar{Z_k})^2,$
$Z(t_i)=U(t_{i+1})-U(t_i)$ and $\bar{Z_k}=\frac{1}{q-1}\sum_{i=1}^{q-1} Z_{(k-1)(q-1)+i}$, $i=1, \ldots, q-1$.
 Also we have that 
 ${\sigma^2_{k,2}}=\hat{\lambda}^{2H}{\sigma_{k-1,2}}^2$, where  $\sigma^2_{k,2}= \mbox{Var}\big(Z(t_{(k-1)(q-1)+i})\big) $. 
Estimating $\sigma_{k,2}^2$ by $S^2_{k,2}$, one can  estimate  the Hurst parameter $H_2$  by 
 $\hat{\mu}_{k,2}=\hat{\lambda}^{2\hat{H}_2}=\frac{S^2_{k,2}}{S^2_{k-1,2 }}$.   
Denoting ${{\mu}}_{k,2}={\hat{\lambda}}_{k}^{2{H_2}}$, $k=2, \ldots, M$, we have the final estimation of $\mu$ as mean of these $\hat{\mu}_{k,2}$.

\subsection{Real data analysis}
As an example of DSI process with drift we consider daily indices of Dow Jones from 25/10/2001 till 28/5/2014 and try to estimate the relevant parameters.
As these indices are no available on Saturdays, Sundays and holidays, the available indices for this duration are 3168 days, which are plotted in Figure 1 (left).
For the duration of the study 6/3/2009 till 14/11/2012 corresponding to the sample points $1800-2600$ we have evaluated a drift line as $g(t)=\hat{a}t+\hat{b}$. So the existence of the drift is clear by the fitted drift line which is shown in Figure 1. The DSI samples which has been evaluated by differencing the data set from this drift line is plotted in the bottom panel of the Figure 1 (right) which shows the DSI behavior.
The fitted red lines reveals the scale intervals of DSI variation with drift for the period 6/3/2009 till 14/11/2012. The end points of these scale intervals are $a_1=1854, a_2=2186, a_3=2466, a_4=2671, a_5= 2785$. The scale parameter is estimated as mean of the ratio of length of successive scale intervals, so the time dependent scale parameters are estimated as $\hat{\lambda}_1 =1.1857, \hat{\lambda}_2 =1.3659, \hat{\lambda}_3 =1.7982$  and their mean  as  $\hat{\lambda}= 1.4499$.

\begin{figure}[h!]
     \begin{center}\label{daj}
     \begin{tabular}{  c  c  }
     \hspace{-.6in}\includegraphics[width=.9\textwidth, height=40mm]{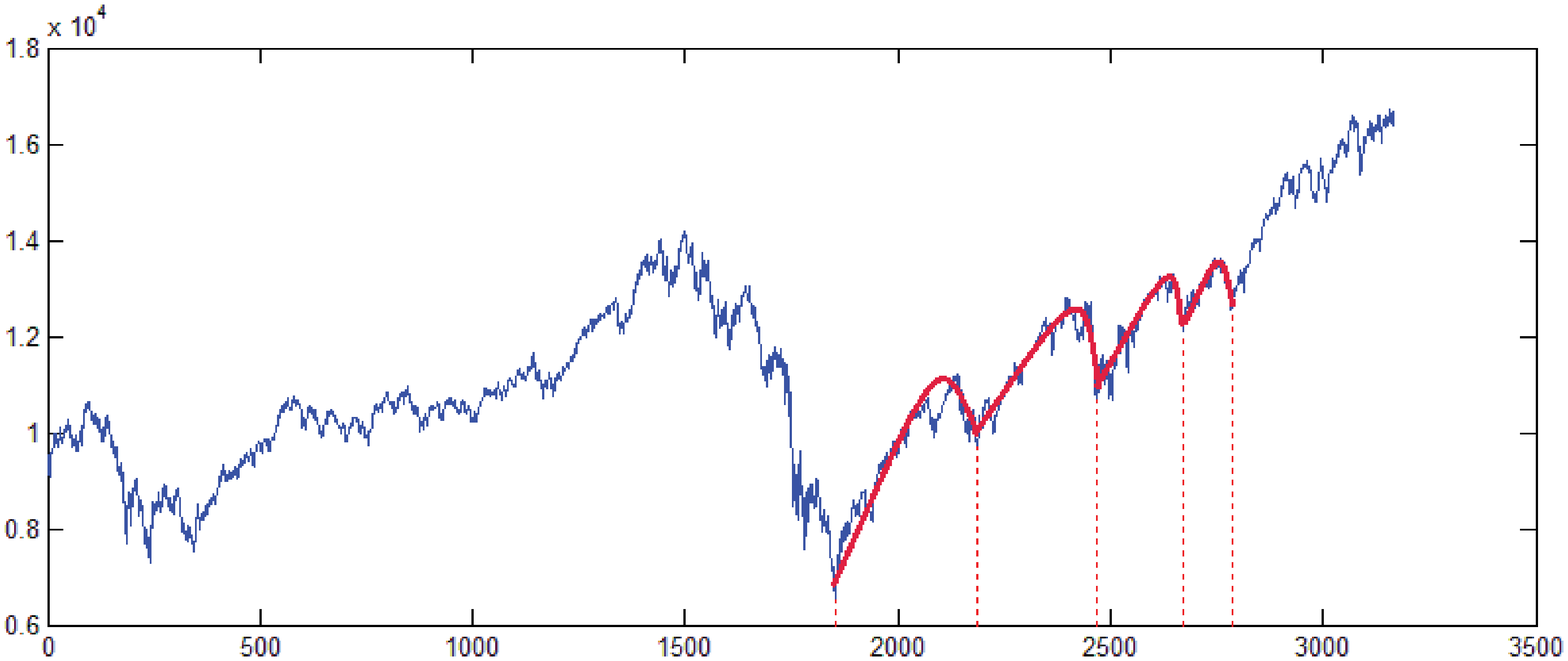}\hspace{-.25in}
      & \\
\hspace{-.55in}\vspace{-0in}
       \includegraphics[width=0.4\textwidth, height=40mm]{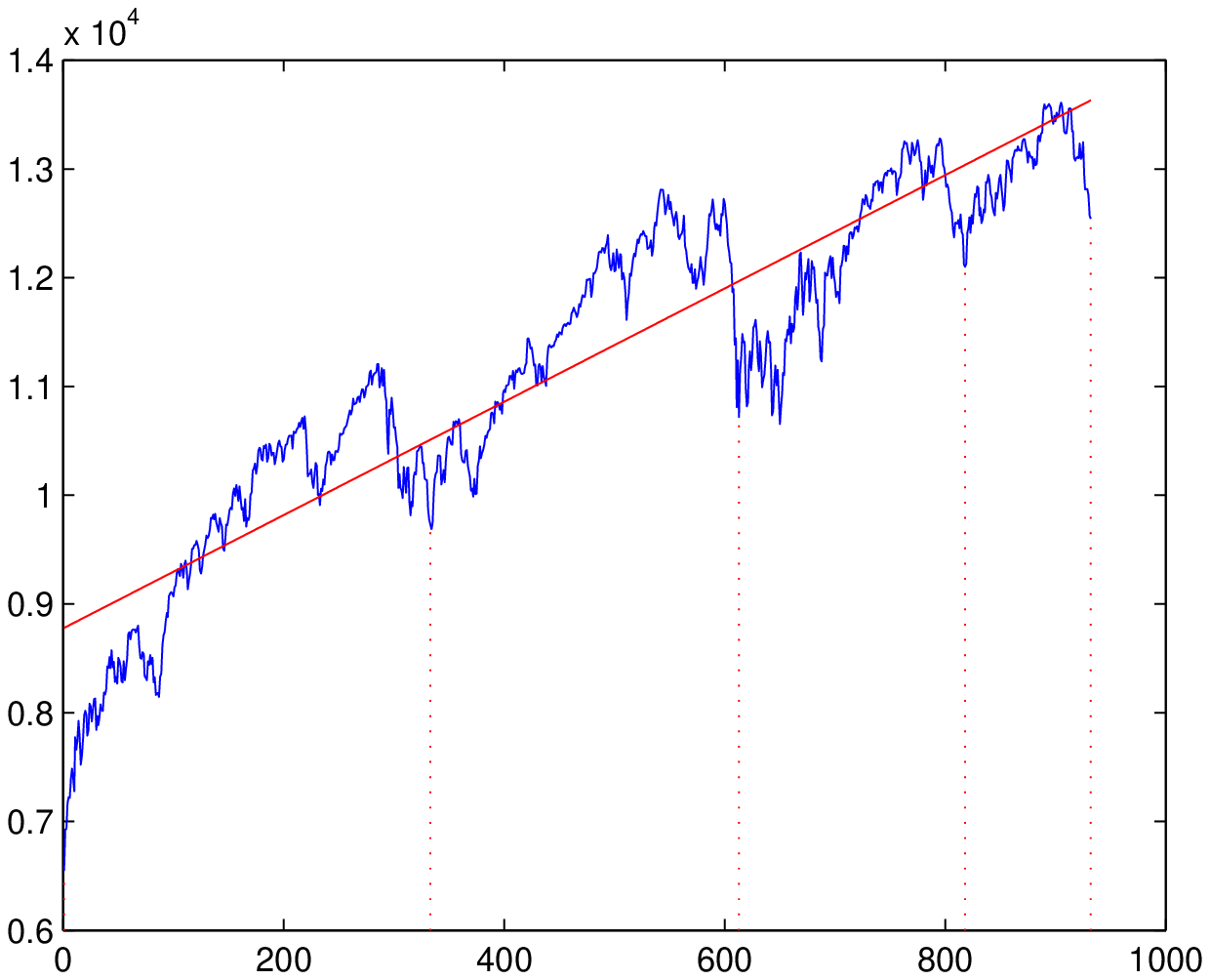}\hspace{-.05in}\hspace{-.1in}\vspace{-0in}
       \includegraphics[width=0.4\textwidth, height=40mm]{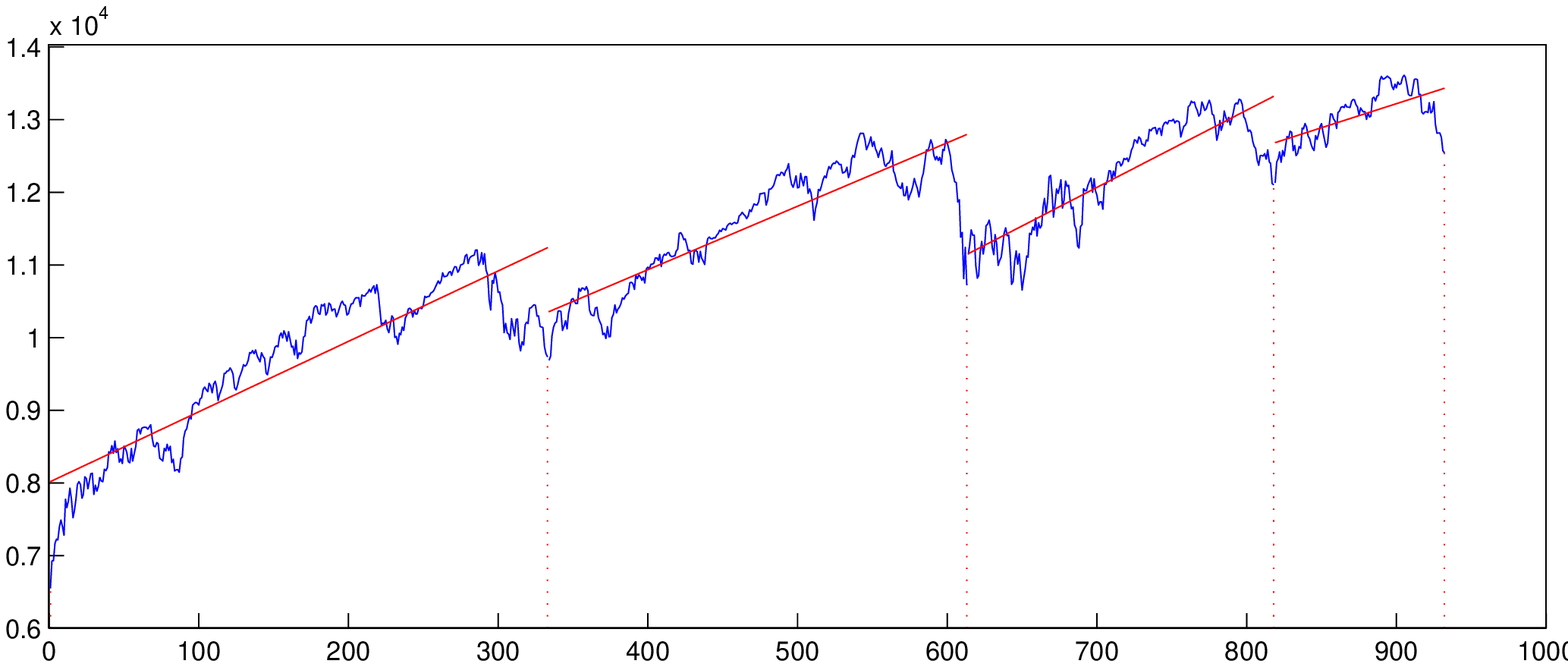}\hspace{-.25in}
          \\
      \end{tabular}
\vspace{0cm}
      \caption{\footnotesize The above figure shows the plot of Dow Jones indices that the DSI behavior is  justified from 6/3/2009 till 14/11/2012. Red lines are borders of  corresponding scale intervals.  The left below figure shows the fitted drift line to the whole duration of DSI behavior, where the slop of the drift line is $5.2$ of share index per unit day. By the right below figure different drifts lines are fitted to the plots of successive scale intervals, where the slop of successive drift lines from the left are $9.7,\; 8.7, \; 10.6$ and $6.6$ respectively. }
      \label{DOWJFig}
      \end{center}
      \end{figure}

\noindent Now we consider some fixed number of equally spaced samples in each two consecutive interval. Then we estimate their corresponding Hurst parameter as the logarithm of the ratio of sample variance of increments in such successive scale intervals.
This is done after eliminating the drifts by different drift lines which have been fitted to the samples of scale intervals separately.  The slop of successive drift lines from the left are $9.7,\; 8.7, \; 10.6$ and $6.6$ share index unit per day, respectively.
Therefore $\log (\hat{\lambda}^{2H_i})=\log (S^2_i/S^2_{i-1})$ where $S^2_i$ is the sample variance of increments inside $i$-th scale interval after eliminating the corresponding drift. Dividing this value by $2\log\hat{\lambda}$ we estimate the time dependent Hurst parameters $H_{i}$ for the variation of $i$-th scale interval with respect to the previous one. A new method for the time dependent Hurst parameter estimation which is more accurate estimation was studied \cite{m114}.
This evaluation leads to the estimation of time dependent Hurst parameters as $H_1=0.5711, H_2=0.1375, H_3=0.8134$ with mean $0.5073$.\\
 We should remind that considering such drift lines has two advantages. First it causes to model the mean changes of the process by such regression lines and the second advantage is that it reveals the true DSI behavior of the process after eliminating such drifts. For more clarification we represents here the estimated Hurst parameters without fitting such drift lines as
$H'_1= 0.7641 , \; H'_2=0.3105$  and $H'_3=1.9404  $ for the Hurst parameters of second, third and fourth scale intervals with respect to the previous scale intervals.   Also when we fit just one drift line to the whole duration of DSI behavior and eliminating such drift from the data the corresponding  Hurst parameters are estimated as $H''_1= 0.9409 , \; H''_2=1.5464$ and $H''_3=0.0567$. Comparing the variations of these estimates with the ones that were estimated by eliminating different drifts for successive scale intervals, reveals that those estimates are more close to each other and are more promising since all estimates are in the range $(0, 1)$.  Also the bottom panel of Figure 1 (right) shows such drift provides much better estimation of share indices.
\\

\section{Comparison of Estimation methods for Hurst parameter of HSSSI process}
In this section we develop  the method  presented in \cite{R-M3} for the  estimation of  Hurst parameter of the scale invariant processes with stationary increments.
We consider the case that the scale invariant processes could have linear drift.
Let $\{X_{t}, t=1,2,\ldots, N\}$ be equally spaced samples of such a process.
	{\color{blue} For this, first we eliminate the drift.}
 We  estimate this drift by the regression line $\hat{X}_t=\hat{a}+\hat{b}t$  by evaluating $\hat{a}$ and $\hat{b}$.
Then we eliminate the drift from the process as $ Z_t=X_t-\hat{X_t}$.
Now we are to estimate the Hurst parameter from this new process $Z_t$. Following the method of sampling of Rezakhah et al. \cite{R-M3}, we consider
sub-samples at points $\{Z_{ik}, i=1,2,\ldots, [N/k]\}$ as the $k$-th sub-sample for some fixed $k\in\mathbb{N}$.
Choosing $k$ depends on the sample size we take  $k\in \{1,...,K^* \}$. We consider  $K^*=\min \{20, N/30$\}.
For every $k\in \{1,...,K^*\}$ we consider two kind of sampled process $\{Z_{i}\}$ and $\{Z_{ik}\}$, where  $i=1,2, \ldots, [N/k]$, and evaluate
first and second order sample variances  for $r=1,2$
\vspace{-5mm}
\begin{equation}
S^2_{r,k,2}=\frac{1}{[\frac{N}{k}]-r}\sum_{i=1}^{[\frac{N}{k}]-r}(Y_{r,i k}-\bar{Y}) ^2,\hspace{.5in} S^2_{r,k,1}=\frac{1}{[\frac{N}{k}]-r}\sum_{i=1}^{[\frac{N}{k}]-r}(Y_{r,i }-\bar{Y})^2 \label{Svar}
\end{equation}
where $Y_{1,j}=Z_{j+1}-Z_j$ and $Y_{2,j}=Z_{j+2}-2Z_{j+1}+Z_j$ correspondingly.
One can easily verify that $Y_{r,ik}\stackrel{d}{=}k^{H'_k}Y_{r,i}$  for $r=1,2$. So $\sigma^2_{r,k,2}=k^{2H'}\sigma^2_{r,k,1}$ ,
where $\sigma^2_{r,k}= \mbox {Var } (Y_{r,ik})$ and  $\sigma^2_{r,1}= \mbox{Var}(Y_{r,i})$.    Thus $S^2_{r,k,2}$ and $S^2_{r,k,1}$ are corresponding sample variances and  estimates of  $\sigma^2_{r,k}$ and $\sigma^2_{r,1}$  respectively.  So
for different values of $k$ we evaluate $\hat{H'}_k$ by the relation.
\vspace{-3mm}
\begin{equation}
\frac{S^2_{r,k,2}}{S^2_{r,k,1}}=k^{2\hat{H'_k}}\label{eq:1},
\end{equation}
and estimate  $H'$  as the mean of such different $\hat{H'_k}$'s  by
\begin{equation}
\hat{H'}= \frac{1}{2(K^*-1)}\sum_{k=2}^{K^*}\log\big(\frac{S^2_{r,k,2}}{S^2_{r,k,1}}\big)/\log(k).\label{Hes}\\
\end{equation}

\noindent Now we compare the performance of the introduced method for estimation of Hurst parameter with FA, DFA and DMA methods.
First we simulate 10000 samples from fBm with different Hurst parameters as $H=0.1, 0.2, \ldots, 0.9$. Then we estimate the Hurst parameters by different methods, FA, DFA, DMA and our two methods first difference (diff 1) and second difference (diff 2) ones with 500 repetitions.  The MSE of the methods are plotted in Figure 2. As it is shown in the figure, the diff 2 method always have much better performance than the previous methods. The diff 1 method is the best for the Hurst parameters between $0.1$ and $0.5$, but for Hurst between $0.6$ and $0.9$ the diff 2 method is the best.

\input{epsf}
\epsfxsize=4.3in
\epsfysize=2.3in
\begin{figure}
\centerline{$\hspace{-.7in}$\epsffile{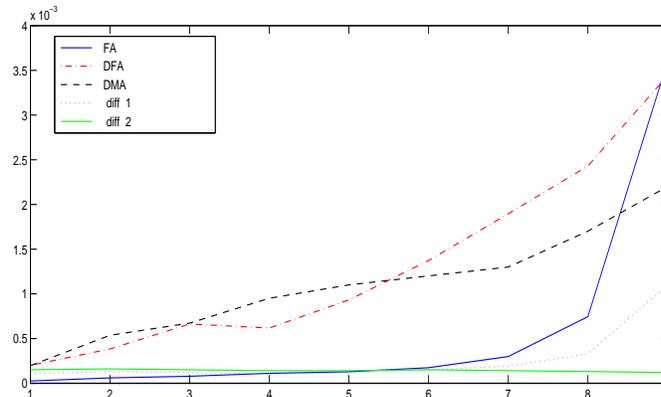}
} 
\vspace{-0.9cm}
\caption{\footnotesize
Mean square error in estimation of Hurst index of using 10000 samples of fBm with 500 repetitions.
}
\end{figure}

\section{Conclusions}
In this paper some heuristic method for the estimation of scale and Hurst parameter of discrete scale invariant processes with piece-wise linear drift.  
 As  many DSI processes are accompanied by some piece-wise linear drift, the method of this paper in estimating  and eliminating of such piecewise drifts, motivates 
further research in this way and provide a platform to extend the applications   
 of DSI processes.   
  So one would expect to have a  better understanding  of the processes involving   DSI behavior.  
  As an example, for the presented part of Dow Jones indices it is shown   that the  DSI behavior is accompany with piece-wise linear drift.  We also presented new method to improve estimation of the Hurst parameter of DSI processes.
 Comparing the presented  method for the estimation of  the Hurst parameter of Hsssi  processes with the celebrated methods   FA, DFA and DMA  shows 
 its superior by producing  less  mean  squared errors. 
This paper could initiate further research and application in compare to the existing methods by applying our method in estimating and eliminating different piece-wise drifts.
Our method  has the privilege to present a better estimation of the Hurst parameter of DSI processes  by eliminating such additional structure for the variations.

\bibliographystyle{amsplain}

\begin{thebibliography}{10}
\bibitem{abr} P. Abry, P. Goncalves, J.L. Vehel (2010) Scale invariance in computer network traffic, {\em Scaling Fractals and Wavelets}, 413-436.
\bibitem{a0} E. Alessio, A. Carbone, G. Castelli, V. Frappietro (2002) Second-order moving average and scaling of stochastic time series, {\em Eur. Phys. J. B 27}, 197.
\bibitem{bal1} G. Balasis, I.A. Daglis, A. Anastasiadis, K. Eftaxias (2011) Detection of dynamical complexity changes in Dst time series using entropy concepts and rescaled range analysis, in "The dynamic magnetosphere", {\em IAGA Special Sopron Book Series}, Springer, Vol. 3, 211-220.
\bibitem{bal2} G. Balasis, I.A. Daglis, C. Papadimitriou, M. Kalimeri, A. Anastasiadis, K. Eftaxias (2009) Investigating dynamical complexity in the magnestosphere using various entropy measures, {\em J. Geophys. Res.}, 114, A00D06.
\bibitem{b-1} G. Balasis, C. Papadimitriou, I.A. Daglis, A. Anastasiadis, L. Athanasopoulou, K. Eftaxias (2011) Signatures of discrete scale invariance in Dst time series, {\em Geophys. Res. Lett.}, 38, L13103/1-6, doi:10.1029/2011GL048019.
\bibitem{b0} M. Bartolozzi, S. Drozdz, D.B. Leieber, J. Speth, A.W. Thomas (2005) Self-similar log-periodic structures in Western stock markets from 2000, {\em International Journal of Modern Physics C}, Vol.16, No.9, pp. 1347-1361.
\bibitem{b2} J. Beran (1994) Satistics for long memory processes, {\em Chapman and Hall, NewYork}.
\bibitem{bor} P. Borgnat, P.O. Amblard, P. Flandrin (2005) Scale invariances and Lamperti transformations for stochastic processes, {\em J. Phys. A: Math. Gen.}, Vol.38, No.10, 2081-2101.
\bibitem{c1} A. Chronopoulouc, C. Tudor, F.G. Viens (2010) Self-similarity parameter estimation and reproduction property for non-Gaussian Hermite processes, {\em Communications in Stochastic Analysis}, Vol. 5, No. 1, 161-185.
\bibitem{c2} J.F. Coeurjolly (2001) Estimating the parameters of a fractional Brownian motion by discrete variations of its sample paths, {\em Statistical Inference for Stochastic Processes}, 4, 199-227.
\bibitem{con} R. Cont, M. Potters, J.P. Bouchaud (1997) Scaling in stock market data: stable laws and beyond, {\em Scalin Invariance and Beyond}, 75-85.
\bibitem{fei} J.A. Feigenbaum, P.G.O. Freund (1996) Discrete scale invariance in stock markets before crashes, {\em Int. J. Mod. Phys.B}, Vol.10, Issue.27, 3737-3745.
\bibitem{hul} J. Hull (2009) Options, Futures, and other Derivatives, 10 ed., Pearson/Prentice Hall.
\bibitem{m11} N. Modarresi, S. Rezakhah (2010), Spectral analysis of Multi-dimensional selfsimilar Markov processes, {\em Journal of Physics A:
    Mathematical and Theoretical}, Vol.43, No.12, 125004, 14pp.
\bibitem{m3} N. Modarresi, S. Rezakhah (2013) A new structure for analyzing discrete scale invariant processes: Covariance and Spectra,
{\em Journal of Statistical Physics}, Vol.152(6), 15pp.
\bibitem{m4} N. Modarresi, S. Rezakhah (2016) Characterization of Discrete Scale Invariant Markov Sequences, {\em Communications in Statistics: Theory and Methods}, 45, 18, 5263-5278 .
\bibitem{p0} C.K. Peng, S.V. Buldyrev, A.L. Goldberger, S. Havlin, F. Sciortino, M. Simons, H.E. Stanley (1992) Long range correlations in nucleotide sequences, {\em Nature 356}, 168.
\bibitem{p00} C.K. Peng, S.V. Buldyrev, S. Havlin, M. Simons, H.E. Stanley, A.L. Goldberger (1994) Moaic organization of DNA nucleotides, {\em Phys. Rev. E 49}, 1685.
\bibitem{m114}  S. Rezakhah, Y. Maleki (2016) Discretization of continuous time discrete scale invariant processes: estimation and spectra,
{\em Journal of Statistical Physics}, Vol.164(2), pp. 438-448.
\bibitem{R-M3} S. Rezakhah, A. Philippe and N. Modarresi (2017) Innovative methods for modeling of scale invariant processes. {\it Communication in Statisctics: Theory and Methods}, Accepted.
\bibitem{rao} R. Rao, S. Lee (2000) Self-similar network traffic characterization through linear scale invariant system models, {\em IEEE Int. Con. On Personal Wireless Communications.} Conference Proceedings.
\bibitem{ros} S.M. Ross (2014) Variations on Brownian Motion, {\em Introduction to Probability Models}, 11th ed., Amsterdam: Elsevier, 612-614.
\bibitem{rud} D.L. Ruderman (1997) Origins of scaling in natural images, {\em Vision Research}, 37, 3385-3398.
\bibitem{sor} D. Sornette (1998) Discrete scale invariance and complex dimensions, {\em Physics Reports}, 297, 239-270.
\bibitem{v1} A. Vidacs, J. Virtamo (1999) ML estimation of the parameters of fBm traffic with geometrical sampling, {\em IFIP TC6, Int. Conf. on Broadband communications 99}, Hong-Kong.
\bibitem{wang} W. Wang, M. Carr, W. Xu, K. Kobbacy (2011) A model for residual life prediction based on Brownian motion with an adaptive drift, {\em Microelectron. Reliab.}, 51, 285-293.
\bibitem{w1} N. Wang, Y. Li, H. Zhang (2010) Hurst exponent estimation based on moving average method, {\em Advances in Wireless Networks and Information Systems}, Vol.72, 137-142.
\bibitem{zho} W.-X. Zhou, D. Sornette (2009) Numerical investigations of descrete scale invariance in fractals and multifractal measures, {\em  Physica A. Stat. Mech. App.}, Vol.388, No. 13, 2623-2639.

\end{thebibliography}

\end{document}